\documentstyle[12pt]{article}

\input epsf.sty
\newdimen\psfigsize
\def\psfigure#1 #2 #3 #4 #5{
    \begin{figure}[tbp]
    \vbox{
    \null\hskip#2\epsfxsize=#1 \epsfbox{#4}
    \vskip 10truept
    \caption {#5 \label{#3}}
    \vskip 0.1truein plus0.2truein}
    \end{figure}
}
\def\psoddfigure#1 #2 #3 #4 #5 #6{
    \begin{figure}[tbhp]
    \vbox{
    \null\hskip#2\epsfxsize=#1 \epsfbox{#5}
    \vskip -#1 \vskip #2 \vskip 10truept
    \vskip 10truept
    \caption {#6 \label{#4}}
    \vskip 0.1truein plus0.2truein}
    \end{figure}
}
\def\figurespace#1 #2 #3 #4 {
    \begin{figure}[tbhp]
    \vbox{
    \psfigsize=#1truein
    \vskip \psfigsize
    \vskip 10truept
    \caption {#4 \label{#3}}
    \vskip 0.1truein plus0.2truein}
    \end{figure}
}
\def\gnufigure#1 #2 #3 #4 #5 #6{
    \begin{figure}[tbhp]
    \vbox{
    \null\hskip#3\epsfxsize=#1 \epsfbox{#5}
    \vskip -#1 \vskip #2 \vskip 10truept
    \vskip 10truept
    \hbox{\null\hskip 1.0in \parbox[t]{4.5in}{ \caption {#6 \label{#4}} } }
    \vskip 0.1truein plus0.2truein}
    \end{figure}
}

\def\etal{{\it et al.}}

\def\Tr{\mathop{\rm Tr}}

\def\Dslash{\mathop{\not\!\! D}}

\def\LP{\left(}		
\def\RP{\right)}	

\def\BE{\begin{equation}}
\def\EE{\end{equation}}
\def\BEA{\begin{eqnarray}}
\def\EEA{\end{eqnarray}}
\def\EL{\nonumber\\}

\newcommand{\la}[1]{\label{#1}}

\newsavebox{\Staple}
\savebox{\Staple}{\begin{picture}(0,0)	
\thicklines
\put(0.0,0.1){\vector(0,1){0.9}}
\put(0.0,1.0){\vector(1,0){0.9}}
\put(0.9,1.0){\vector(0,-1){0.9}}
\end{picture}}

\newsavebox{\FiveStaple}
\savebox{\FiveStaple}{\begin{picture}(0,0)	
\thicklines
\put(0.0,0.1){\vector(0,1){0.9}}
\put(0.0,1.0){\vector(1,1){0.5}}
\put(0.5,1.5){\vector(1,0){0.9}}
\put(1.4,1.5){\vector(-1,-1){0.5}}
\put(0.9,1.0){\vector(0,-1){0.9}}
\end{picture}}

\newsavebox{\SevenStaple}
\savebox{\SevenStaple}{\begin{picture}(0,0)	
\thicklines
\put(0.0,0.1){\vector(0,1){0.9}}
\put(0.0,1.0){\vector(1,1){0.5}}
\put(0.5,1.5){\vector(1,2){0.3}}
\put(0.8,2.1){\vector(1,0){0.9}}
\put(1.7,2.1){\vector(-1,-2){0.3}}
\put(1.4,1.5){\vector(-1,-1){0.5}}
\put(0.9,1.0){\vector(0,-1){0.9}}
\end{picture}}

\newsavebox{\LepageStaple}
\savebox{\LepageStaple}{\begin{picture}(0,0)	
\thicklines
\put(0.0,0.1){\vector(0,1){0.9}}
\put(0.0,1.0){\vector(0,1){0.9}}
\put(0.0,1.9){\vector(1,0){0.9}}
\put(0.9,1.9){\vector(0,-1){0.9}}
\put(0.9,1.0){\vector(0,-1){0.9}}
\end{picture}}

\newsavebox{\Link}
\savebox{\Link}{\begin{picture}(0,0)	
\thicklines
\put(0.0,0.0){\vector(1,0){0.9}}
\end{picture}}

\newsavebox{\Naik}
\savebox{\Naik}{\begin{picture}(0,0)	
\thicklines
\put(0.0,0.0){\vector(1,0){0.9}}
\put(1.0,0.0){\vector(1,0){0.9}}
\put(2.0,0.0){\vector(1,0){0.9}}
\end{picture}}

\newsavebox{\OneFatNaik}
\savebox{\OneFatNaik}{\begin{picture}(0,0)	
\thicklines
\put(0,0){\usebox{\Link}\makebox(0,0)}
\put(1.1,0)+
\put(1.5,0){\usebox{\Naik}\makebox(0,0)}
\put(4.6,0)+
\put(5.1,0){\usebox{\Staple}\makebox(0,0)}
\end{picture}}

\begin{document}

\begin{titlepage}
\baselineskip=16pt
\rightline{\bf hep-lat/9903032}
\baselineskip=20pt plus 1pt
\vspace{1.5cm}

\centerline{\Large \bf Variants of fattening and flavor symmetry
restoration}
\bigskip
\centerline{{\bf Kostas~Orginos} {\it and} {\bf Doug Toussaint} }
\centerline{\it
Department of Physics, University of Arizona, Tucson, AZ 85721, USA}
\bigskip
\centerline{\bf R.L.~Sugar }
\centerline{\it
Department of Physics, University of California, Santa Barbara, CA 93106, USA}
\vskip 0.5in
\centerline{ The MILC Collaboration }

\narrower
We study the effects of different ``fat link'' actions for Kogut-Susskind
quarks on flavor symmetry breaking.  Our method is mostly empirical -
we compute the pion spectrum with different valence quark actions on
common sets of sample lattices.
Different actions are compared, as best we can, at equivalent physical
points.
We find significant reductions in
flavor symmetry breaking relative to the conventional or to the ``link
plus staple'' actions, with a reasonable cost in computer time.
We also develop and test a scheme for
approximate unitarization of the fat links.
While our tests have concentrated on the valence quark action, 
our results will be useful in designing simulations with
dynamical quarks.
\end{titlepage}

\section{Introduction}

As lattice QCD computations evolve towards quantitative predictions, it
becomes increasingly important to make better use of computational
resources by using more sophisticated discret\-izations of the continuum
action, or ``improved actions''.  With Kogut-Susskind quarks the errors
from the lattice discretization of the fermion action
are proportional to $a^2$, where $a$ is the lattice spacing.
In contrast, for the (unimproved) Wilson quark formulation, the errors
are proportional to $a^1 g$.  However, for the lattice spacings that
are now practical for calculations with dynamical quarks, the effects
of these errors in the Kogut-Susskind formulation are not small.  In
particular, the breaking of flavor symmetry is a large effect.
The Kogut-Susskind formulation naturally describes four flavors of
quarks, which is conventionally reduced to two in dynamical simulations
by taking the square root of the determinant.  However, only a U(1)
subgroup of the original SU(4) chiral symmetry is an exact symmetry, 
so for nonzero lattice spacing only one of the sixteen pseudoscalar
mesons has a vanishing mass when the quark mass vanishes.
In principle, flavor symmetry breaking is also present for vector
mesons, nucleons, etc., but these effects are generally small, and
we can concentrate on the pseudoscalar mesons.

Flavor symmetry breaking can be reduced by ``fattening'' the links,
which means that the conventional parallel transport using the link
matrix is replaced by an average over paths connecting the points.
This was demonstrated in Ref.~\cite{MILC_FATLINKS}, where the single
link was replaced by an average of the simple link and three link
paths, or ``staples''.  This modification was introduced as the
simplest possible gauge invariant modification of the nearest neighbor
coupling, and the relative weight of the staples and simple connection
was treated as a free parameter.  The improvement of flavor symmetry
was found to be fairly insensitive to the value of the weighting
coefficient.  Lepage pointed out that this improvement could be
understood as an introduction of a form factor suppressing the
coupling to high momentum gluons which scatter quarks from one corner
of the Brillouin zone to another\cite{LEPAGE_TSUKUBA}.  Lag\"ae and
Sinclair used this understanding to construct an action where
successive smearings in all four directions canceled the tree level
couplings to all gluons with any momentum component equal to $\pi/a$,
and showed that this further reduced flavor symmetry
breaking\cite{ILLINOIS_FAT}.  This action includes paths to nearest
neighbor points that are up to nine links in length.  A fat link
Kogut-Susskind fermion action, motivated by a perfect Kogut-Susskind
fermion action, was tested by Bietenholz and Dilger\cite{WB_HD} on the
2D Schwinger Model. They found that their action shows very good
scaling for the pion and the eta masses. Furthermore, perfect action
motivated fattening has been used by DeGrand\cite{TD99} for Wilson-like
fermions, showing very good scaling and renormalization factors very
close to unity. In an earlier paper we investigated the single staple
fattening, together with the Naik term\cite{NAIK} improving the dispersion
relation in more detail\cite{US_PAPER1}.  Ref.~\cite{US_PAPER1} also
describes an algorithm for using these actions in full QCD
simulations, and extends studies of flavor symmetry breaking to the
nonlocal pions, which turn out to have much larger breakings than the
local non-Goldstone pion which was used in the earlier studies.

This paper extends our studies of flavor symmetry breaking to actions
involving paths longer than three links.  Many of these results were
briefly presented at the Lattice-98 conference\cite{US_LAT98}.  We
investigate actions that involve five and seven link paths to the
nearest neighbor point, in addition to the one and three link paths.
The coefficients of the various paths are chosen to minimize or to
completely eliminate the tree level couplings to gluons with
transverse momentum components $\pi/a$.  Motivated by encouraging
results using links fattened by ``Ape smearing''\cite{TD_AH_TK}, we
also try a variant of the fattened action which makes the fattened
links approximately unitary.

Recently Lepage has analyzed flavor symmetry breaking using the
language of Symanzik improvement\cite{LEPAGE98}.  In addition to
providing a simple construction of the seven-link action which cancels
tree level couplings at momentum $\pi$, Lepage points out in this work
that the fattening which reduces flavor symmetry breaking introduces
an error of order $a^2$ in the low momentum physics (effectively
an $a^2p^2$ error, where $p$ is a momentum), and shows how
this error can be removed by an additional term in the action.

This work concentrates on the effect of the valence quark action on the
flavor symmetry breaking.  We have not done a study of the effects of
changing the dynamical quark action; however, we have found in previous 
studies that improvements in flavor symmetry arising from changes in the 
valence quark action are indicative 
of improvements in full QCD simulations\cite{US_PAPER1}. 
Also, we have not studied
scaling, or the independence of mass ratios on the lattice spacing,
although we expect that they will show improved scaling properties.
Our concentration on the splittings of the pions is motivated by
experience that suggests that this is the worst practical problem with
Kogut-Susskind quarks.  (For example, in Ref.~\cite{US_PAPER1} we found
that rotational symmetry of the pion propagator was essentially
restored at coarser lattice spacings than was the flavor symmetry.)
Furthermore, we expect that the study of the pion spectrum provides a   
good guide to the quality of actions, since the                      
low energy dynamics of QCD is approximately the dynamics of a pion gas.

\section{The actions}

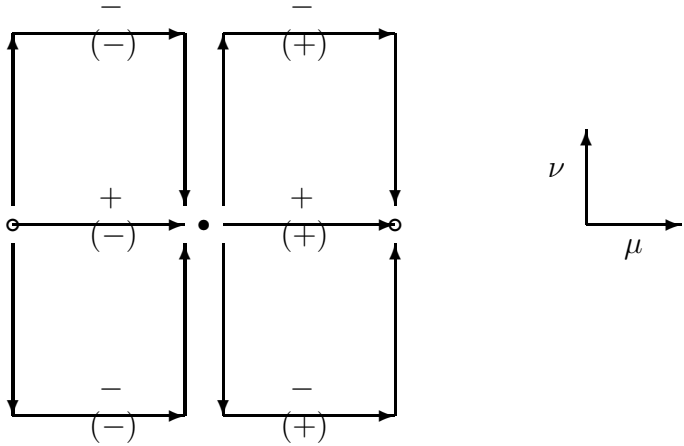
\begin{figure}[tbh]
\setlength{\unitlength}{1.0in}
\begin{center}\begin{picture}(6.0,2.0)(-2.0,-1.0)
\thicklines
\put( 2.0, 0.0){\vector(1,0){0.5}}
\put( 2.0, 0.0){\vector(0,1){0.5}}
\put( 2.2,-0.15){$\mu$}
\put( 1.8, 0.25){$\nu$}

\put( 0.0, 0.0){\circle*{0.05}}
\put( 1.0, 0.0){\circle{0.05}}
\put(-1.0, 0.0){\circle{0.05}}

\put( 0.1, 0.0){\vector(1,0){0.9}}
\put(-1.0, 0.0){\vector(1,0){0.9}}

\put( 0.1, 0.1){\vector(0,1){0.9}}
\put( 0.1, 1.0){\vector(1,0){0.9}}
\put( 1.0, 1.0){\vector(0,-1){0.9}}

\put(-1.0, 0.1){\vector(0,1){0.9}}
\put(-1.0, 1.0){\vector(1,0){0.9}}
\put(-0.1, 1.0){\vector(0,-1){0.9}}

\put( 0.1,-0.1){\vector(0,-1){0.9}}
\put( 0.1,-1.0){\vector(1,0){0.9}}
\put( 1.0,-1.0){\vector(0,1){0.9}}

\put(-1.0,-0.1){\vector(0,-1){0.9}}
\put(-1.0,-1.0){\vector(1,0){0.9}}
\put(-0.1,-1.0){\vector(0,1){0.9}}

\thinlines
\put( 0.45, 0.1){+}
\put( 0.40,-0.1){(+)}
\put( 0.45, 1.1){$-$}
\put( 0.40, 0.9){(+)}
\put( 0.45,-0.9){$-$}
\put( 0.40,-1.1){(+)}

\put(-0.55, 0.1){+}
\put(-0.60,-0.1){($-$)}
\put(-0.55, 1.1){$-$}
\put(-0.60, 0.9){($-$)}
\put(-0.55,-0.9){$-$}
\put(-0.60,-1.1){($-$)}

\end{picture}\end{center}
\caption{
  \label{PLANEFIGURE}
  A simple link and staples in one plane, showing the relative signs
of the coupling to a gluon with momentum component $\pi$ in the
transverse direction (unparenthesized), and a gluon with momentum
$\pi$ in the longitudinal direction (parenthesized).
}
\end{figure}

\begin{figure}[tbh]
\setlength{\unitlength}{1.0in}
\begin{center}\begin{picture}(6.0,2.0)
\put(0.5,0){\usebox{\Link}\makebox(0,0)}
\put(1.5,0){\usebox{\Staple}\makebox(0,0)}
\put(2.5,0){\usebox{\FiveStaple}\makebox(0,0)}
\put(3.5,0){\usebox{\SevenStaple}\makebox(0,0)}
\put(5.0,0){\usebox{\LepageStaple}\makebox(0,0)}
\end{picture}\end{center}
\caption{
   \label{PATHSET}
   The simple link, three link staple, five link staple and seven link
   staple used in suppressing flavor symmetry breaking.  The final
   path is the five link path used, following Lepage, to correct the
   small momentum form factor.
}
\end{figure}
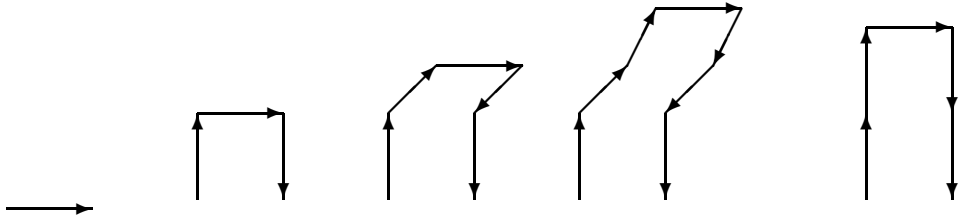

Figure \ref{PLANEFIGURE} illustrates the coupling to gluons with
momentum components $\pm \pi$.  The figure illustrates the single
link couplings in $D_\mu$, connecting the central point to both the
forwards and backwards nearest neighbors, as well as three link staples
connecting to the same points.  Each horizontal arrow represents a parallel
transport by $U_\mu \approx {\bf 1} + iga A_\mu$.  The directions of
the arrows in the coupling to the backwards neighbor include one minus
sign because this transport involves $U^\dagger$ and another minus sign
appearing explicitly in the derivative.  Now consider the coupling to
a $\mu$-direction gluon ($A_\mu(\pi \hat e_\nu)$) with momentum component $\pm \pi$ in
the $\nu$ direction.  The $\pm$ signs above the $\mu$ direction links
indicate the sign of the coupling of this gluon relative to the forward
simple link.  Thus, if $c_1$ and $c_3$ are the weights of the one-link
and three-link paths, the coupling to this gluon from the paths pictured
here is $c_1-2c_3$.  (This is not the whole story --- one must also  
include the staples in the directions that are not shown in this
figure.)
Similarly, the parenthesized signs below the $\mu$ direction links show
the relative coupling to a gluon with momentum $\pi$ in the $\mu$
direction ($A_\mu(\pi \hat e_\mu)$).  Note that this coupling is
automatically cancelled between  the forward and backward parts of
$D_\mu$, so we don't have to worry about longitudinal momentum $\pi$.
Now one might worry that the $\nu$ direction links, required to keep
the expression gauge invariant, might introduce couplings to $\nu$
direction gluons with momentum components $\pm \pi$. But
for $A_\nu(\pi \hat e_\mu)$ gluons, the contributions of the vertical
links in the center cancel, and the contribution of the links from
the left and right sides cancel, since they are separated by $2a$ and
traversed in opposite directions.
Similarly, the coupling to $A_\nu(\pi \hat e_\nu)$ gluons cancels
between the top and bottom halves of the figure.
This argument extends to the five and seven link paths illustrated in
Fig.~\ref{PATHSET}, so the end result is that we can compute the
couplings to gluons with momentum components equal to $\pi$ by just
considering the $\mu$ direction links in $D_\mu$.  Taking into account
the multiplicities of the various paths and using $C_n$ for the weight
of the $n$ link paths, the couplings are given below.  In these
expressions the coefficients explicitly show how many paths of each
length contribute with positive weight and how many with negative
weight.
We also write explicitly $(2 C_5)$ and $(6 C_7)$ to indicate that
in $\Dslash_x$
there are two shortest paths connecting the starting point
to the $\hat x$ direction link displaced by $+\hat y + \hat z$,
and six paths to the
link displaced by $+\hat y + \hat z + \hat t$.
\begin{itemize}

\item{
Coupling to $k = (0,0,0,0)$:  $C_1 + 6 C_3 + 12 (2 C_5) + 8 (6 C_7)$
}

\item{
Coupling to $k = (0,\pi,0,0)$:  $C_1 + (4-2)C_3 + (4-8)(2 C_5) + (-8)(6 C_7)$
}

\item{
Coupling to $k = (0,\pi,\pi,0)$:  $C_1 + (2-4)C_3 + (4-8)(2 C_5) + (+8)(6 C_7)$
}

\item{
Coupling to $k = (0,\pi,\pi,\pi)$:  $C_1 + (-6)C_3 + (12)(2 C_5) + (-8)(6 C_7)$
}

\end{itemize}

If we are willing to use all the paths up to length seven, we can
normalize the zero momentum coupling to one and set all the others to
zero with $C_1=1/8$, $C_3=1/16$, $(2 C_5)=1/32$ and $(6 C_7)=1/64$.  This
defines our ``Fat7'' action and, with tadpole improved coefficients,
the ``Fat7tad'' action.
However, it is interesting to ask how helpful this complexity is ---
could we get by with shorter paths?
If we restrict ourselves to five link and shorter paths, we can no
longer satisfy all of these conditions.  However, we can choose the
couplings to minimize the maximum of the couplings to the high momentum
gluons.  This leads to $C_1=1/7$, $C_3=1/14$ and $(2 C_5)=1/28$, with
all couplings to high momentum gluons reduced by a factor of seven.
This defines our ``Fat5'' action.

Here we note that in two dimensions the equivalent of
our ``Fat7'' action has only a 3-link staple. Our tree level
formula for the relative weight of the staple would be 0.25 in two
dimensions.
This is very close to the relative weight of 0.238 introduced in the
approximate perfect action constructed in Ref.\cite{WB_HD} for the 2D
Schwinger Model.

All of these actions can be tadpole improved by inserting a factor
of $(1/u_0)^{L-1}$ in the coefficient of each path, where $L$ is
the length of the path.  Using $L-1$ instead of $L$ amounts to absorbing
one power of $u_0$ into the quark mass.  We use the average plaquette
to define $u_0$.

While the paths shown in Fig.~\ref{PATHSET} can be used to reduce or
eliminate couplings to gluons with transverse momentum components
equal to $\pi$, Lepage has pointed out that they have the undesirable
effect of modifying the coupling to gluons with small nonzero transverse
momentum components, essentially by introducing a second derivative
coupling proportional to $a^2 p^2$\cite{LEPAGE98}.  Following
Ref.~\cite{LEPAGE98}, we can correct for this by introducing a
flavor conserving five link path, $ +\hat y +\hat y +\hat x -\hat y
-\hat y$, into $D_x$, giving an action with no tree level order $a^2$
corrections.  While it is not clear what observable quantities are
affected by this correction, it is a relatively cheap and aesthetically
pleasing addition to the Fat7+Naik action.
We have tested flavor symmetry breaking with this action, which we
call the ``Asq'' action, or, when the coefficients are tadpole
improved, the ``Asqtad'' action.

To summarize and clarify these actions, we give the coefficients
of the paths in the $a^2$ improved action in a form useful for
simulation.  Here $c_1$ is the coefficient of the simple link.
$c_3$, $c_5$ and $c_7$ are the coefficients of the three, five and
seven link paths in Fig.~\ref{PATHSET}.  $c_N$ is the coefficient
of the three link path to the third nearest neighbor (Naik term), and
$c_L$ is the coefficient of the five link path implementing the
correction introduced by Lepage.  The origin of each term in the
coefficients is identified by a subscrpt: ``F'' for flavor symmetry,
``N'' for the Naik correction to the quark dispersion relation, and ``L''
for the small momentum form factor correction.  The factors of $1/2$ and
$1/6$ in $c_5$ and $c_7$ compensate for the number of different paths
connecting the starting point to the $\mu$ direction link parallel
to the simple link.  For example, the $x$ direction link displaced by
$+\hat y +\hat z$ (coefficient $c_5$) is included in both
$+\hat y\  +\hat z\  +\hat x\  -\hat z\  -\hat y$ and
$+\hat z\  +\hat y\  +\hat x\  -\hat y\  -\hat z$, and we average over both
paths.
\BEA
       c_1 &=& ( 1/8)_F +(3/8)_L+(1/8)_N \EL        
       c_3 &=& ( 1/16)_F \EL 	            
       c_5 &=& ( 1/32)_F(1/2) \EL          
       c_7 &=& ( 1/64)_F(1/6) \EL   
       c_L &=& (-1/16 )_L \EL                  
       c_N &=& (-1/24)_N 	            
\EEA

We have also experimented with an action in which the fat links are
approximately unitary.  Theoretically, it is not clear why unitarity
should be a concern in suppressing flavor symmetry breaking.  However,
results in Ref.~\cite{TD_AH_TK} led us to consider ``APE-smeared''
links, where an average over the single link and nearby paths is
constructed as above, and then the resulting fat link is projected
back onto the nearest element of SU(3).  Of course, such actions
are not explicitly expressed as a sum over paths, and dynamical
simulations using them would require an extension of our algorithms.
However, spectrum calculations using APE-smeared links for the valence
quarks are straightforward.  In particular, we ran a calculation
using a single iteration of APE smearing on a fairly coarse lattice.
This action differs from the one link plus staple action only in the
projection onto a special unitary matrix, yet, as will be seen
in a later section, it produced smaller flavor symmetry breaking
than the one link plus staple plus Naik action.
(The Naik term has little effect on flavor symmetry.)
This motivated us to construct an action for which the fat links are 
approximately unitary, but are still expressed as an explicit sum
over paths.

An almost unitary matrix $M$ can be expressed as a unitary matrix $U$ times
a correction:
\BE M = U \LP {\bf 1} + \epsilon \RP \la{UDEF}\ \ ,\EE
where $\epsilon$ is Hermitian.
Now look at $M^\dagger M$ to first order in $\epsilon$
\BE M^\dagger M 
= \LP {\bf 1} + \epsilon \RP U^\dagger U \LP {\bf 1} + \epsilon \RP
\approx {\bf 1} + 2\epsilon\ \ .
\EE
Now, inverting Eq.~\ref{UDEF} to first order, 
\BE U = M\LP {\bf 1}-\epsilon\RP = \frac{3}{2} M - \frac{1}{2}M M^\dagger M 
\ \ .\EE
We want to use this equation to make a fat link approximately unitary.
Let $M$ be a generic fat link:
\BE M = L + \alpha S \EE
where $L$ is the simple link and $S$ is some sum over other paths
connecting the ends of $L$.  $S$ may be the sum of the staples for
a simple fattening, or something more complicated involving longer
paths.
We will work to first order in $\alpha$.  Customarily we also rescale
$M$ by some factor which is $1+b\alpha$.  However, this factor cancels
in the derivation, so we suppress it here.
Unitarize $M$ as above to make an approximately
unitary fat link $F$:
\BEA 
 F &=&\frac{3}{2} M - \frac{1}{2}M M^\dagger M \EL
 &=& \frac{3}{2} \LP L + \alpha S \RP
 - \frac{1}{2} \LP L+\alpha S\RP\LP L^\dagger + \alpha S^\dagger \RP
  \LP L+\alpha S\RP
\EEA
Expand, keeping terms up to order $\alpha$ and using $L^\dagger L = {\bf
1}$, and remarkable cancellations occur
\BE F = L + \frac{\alpha}{2}\LP S - L S^\dagger L \RP \la{FDEF}\EE

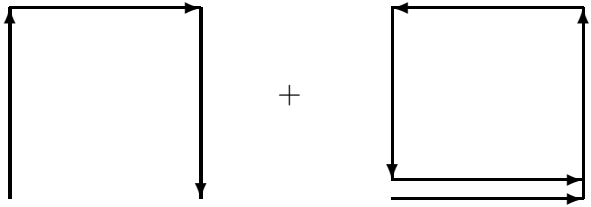
\begin{figure}[tb]
\setlength{\unitlength}{1.0in}
\begin{center}\begin{picture}(6.0,2.0)
\thicklines

\put(1.0, 0.0){\vector(0,1){1.0}}
\put(1.0, 1.0){\vector(1,0){1.0  }}
\put(2.0, 1.0){\vector(0,-1){1.0}}

\put(2.4, 0.5){+}

\put(3.0, 0.0){\vector(1,0){1.0}}
\put(4.0, 0.0){\vector(0,1){1.0}}
\put(4.0, 1.0){\vector(-1,0){1.0}}
\put(3.0, 1.0){\vector(0,-1){0.9}}
\put(3.0, 0.1){\vector(1,0){1.0}}
\end{picture}\end{center}
\caption{
   \label{UNITARIZED}
   An approximately unitary variant of the three link staple.
}
\end{figure}

$F$ is an approximately unitary fat link expressed as a sum over
paths --- a form suitable for dynamical simulations.
The only change is that the sum over fattening paths, $S$, has been
replaced by an average of the paths traversed in each direction,
$\frac{1}{2}\LP S-S^\dagger \RP$, with $L$ inserted as necessary to
maintain gauge invariance.
At this point one can forget about how it was derived, and just verify
from Eq.~\ref{FDEF} that $F^\dagger F = {\bf 1}$ to order $\alpha$.
It is also easy to verify, by setting link matrices to ${\bf 1} + i A_\mu$,
that $F$ averages $A_\mu$ over position in the same way that $M$
did.  The minus sign in front of the $L S^\dagger L$ term compensates
for the fact that this path is now traversed in the opposite direction.
This action is illustrated in Fig.~\ref{UNITARIZED} for the case where
$S$ is the three link staple.

The two terms can be made to look more symmetric by factoring $L$
out on the right (or left).
\BE F = \LP {\bf 1} + \frac{\alpha}{2}\LP S L^\dagger - L S^\dagger \RP
\RP L \la{FDEF2}\EE
In this form, we are just following the parallel transport by $L$
with a difference of going around closed loops in opposite directions,
or a term proportional to $F_{\mu\nu}$.

\section{The simulations}

Most of our spectrum calculations used two of the same sets of sample
lattices used in Ref.~\cite{US_PAPER1}.
These were lattices with two
flavors of dynamical quarks, where the dynamical quarks used the
"Staple+Naik" action, and a Symanzik improved gauge action
at $10/g_{imp}^2=7.3$ and $7.5$.
At $10/g_{imp}^2=7.3$ we used $12^3\times 32$ lattices with dynamical quark
masses of $0.02$ and $0.04$, with lattice spacings determined from $r_0$
\cite{SOMMER} of about 0.15 fm. and 0.16 fm. respectively. 
The $10/g_{imp}^2=7.5$ runs used $16^3\times 48$ lattices with $m_q=0.015$
and $0.030$, with lattice spacings of $0.13$ fm and $0.14$ fm,
respectively.
Sample sizes ranged from 48 to 60 lattices\cite{US_PAPER1}, with four
source time slices per lattice for spectrum calculations.
In addition
we computed the meson spectrum on a set of eleven large ($32^3\times 64$)
quenched lattices with a lattice spacing around $0.07$ fm.,
using the one plaquette gauge action at $6/g_{conv}^2=6.15$.

For each of the three gauge couplings we computed spectra at two values
of the quark mass, which allows us to interpolate results to make fair
comparisons of the actions.  (For the two flavor runs, the sea quark
mass was also changed.)  The meson masses were interpolated assuming
that the squared meson masses are linear in the quark mass.
One approach to comparing actions is to interpolate to a fixed
ratio of $m_G$ to $m_\rho$, where $m_G$ is the Goldstone pion mass.
In so doing, we are letting each action ``determine its own lattice
spacing''.  Since the $\rho$ mass is also dependent on the valence
quark action, this results in a slightly different estimate of the lattice
spacing for each action.  An alternate approach is to assume that each
set of lattices has a fixed lattice spacing, which could be determined
either from the rho mass for a fixed choice of action, or from the
static quark potential.  Operationally, this results in interpolating
all the valence actions on a set of lattices to the same pion mass,
and we choose the Goldstone pion for this interpolation.
For the runs with $10/g_{imp}^2=7.3$ and $7.5$
we will present results with both sets of assumptions.  For the fine
lattices, with $6/g_{conv}^2=6.15$, the differences in the $\rho$
masses with different actions are insignificant, which is expected
as the lattice spacing gets smaller.

As in Ref.~\cite{US_PAPER1}, we parameterize the splitting of the pions by the
dimensionless quantity 
\BE 
\delta_2 = \frac{m_\pi^2 -m_G^2}{m_\rho^2-m_G^2} \ \ \ ,        
\EE 
where $m_\pi$ is one of the
non-Goldstone pion masses, $m_G$ is the Goldstone pion mass, and      
$m_\rho$ is one of the local $\rho$ masses.  Since the $\rho$ masses are 
nearly degenerate, it makes little difference which one we use.  
In our analysis we used the local $\gamma_i \otimes \gamma_i$ ($\rho_2$)
mass because for heavy quarks it is often
estimated more accurately.  
We use the squared meson masses since they are
approximately linear in the quark mass and we
interpolate and extrapolate in quark mass.
Empirically $\delta_2$ is
fairly insensitive to the quark mass, and the theoretical analysis
in Ref.~\cite{SHARPE} predicts that the numerator is independent
of quark mass for small quark masses.
The squared Goldstone pion mass is
subtracted in the denominator to give more sensible behavior at large
quark mass.  At large $am_q$, all of the meson masses become  
degenerate, but it is still possible to use the difference between the
Goldstone pion mass and the rho mass as a natural scale for flavor
symmetry breaking.

\renewcommand{\arraystretch}{1.5}
\begin{center}\begin{table}
\label{ACTIONTABLE}
\caption{Path coefficients for the various quark actions.  The first column is a name for the
action.  The remaining columns are the coefficients for the paths in Fig.~\protect\ref{PATHSET}
and for the connection to the site at distance three (Naik term).
Notice that in the tadpole improved actions we have absorbed one power of $u_0$ into the quark
mass, so $u_0$ appears to the power ${\rm length}-1$.
}
\begin{tabular}{lcccccc}
Action		& Link			& 3-staple 		& 5-staple
		& 7-staple 		& Lepage 		& Naik \\
OneLink		& 1			& -			& -
		& -			& -			& -	\\
Staple+Naik	& $\frac{9}{8}\,\frac{1}{4}$& $\frac{9}{8}\,\frac{1}{8}$& -	
		& -			& -			& $\frac{9}{8}\,\frac{-1}{27}$	\\
Fat5		& $\frac{1}{7}$		& $\frac{1}{7}\,\frac{1}{2}$& $\frac{1}{7}\,\frac{1}{8}$	
		& -			& -			& -	\\
Fat5tad		& $\frac{1}{7}$		& $\frac{1}{7}\,\frac{1}{2}u_0^{-2}$& $\frac{1}{7}\,\frac{1}{8}u_0^{-4}$
		& -			& -			& -	\\
Fat7		& $\frac{1}{8}$		& $\frac{1}{8}\,\frac{1}{2}$& $\frac{1}{8}\,\frac{1}{8}$	
		&$\frac{1}{8}\,\frac{1}{48}$& -			& -	\\
Fat7tad		& $\frac{1}{8}$		& $\frac{1}{8}\,\frac{1}{2}u_0^{-2}$& $\frac{1}{8}\,\frac{1}{8}u_0^{-4}$
		&$\frac{1}{8}\,\frac{1}{48}u_0^{-6}$& -			& -	\\
Asqtad		& $\frac{1}{8}+\frac{3}{8}+\frac{1}{8}$		& $\frac{1}{8}\,\frac{1}{2}u_0^{-2}$& $\frac{1}{8}\,\frac{1}{8}u_0^{-4}$
		&$\frac{1}{8}\,\frac{1}{48}u_0^{-6}$&$\frac{-1}{16}u_0^{-4}$
		&$\frac{-1}{24}u_0^{-2}$	\\
\end{tabular}
\end{table}\end{center}
\renewcommand{\arraystretch}{1.0}

\section{Results}

\begin{figure}[tb]
\epsfxsize=6.0in
\epsfysize=6.0in
\epsfbox{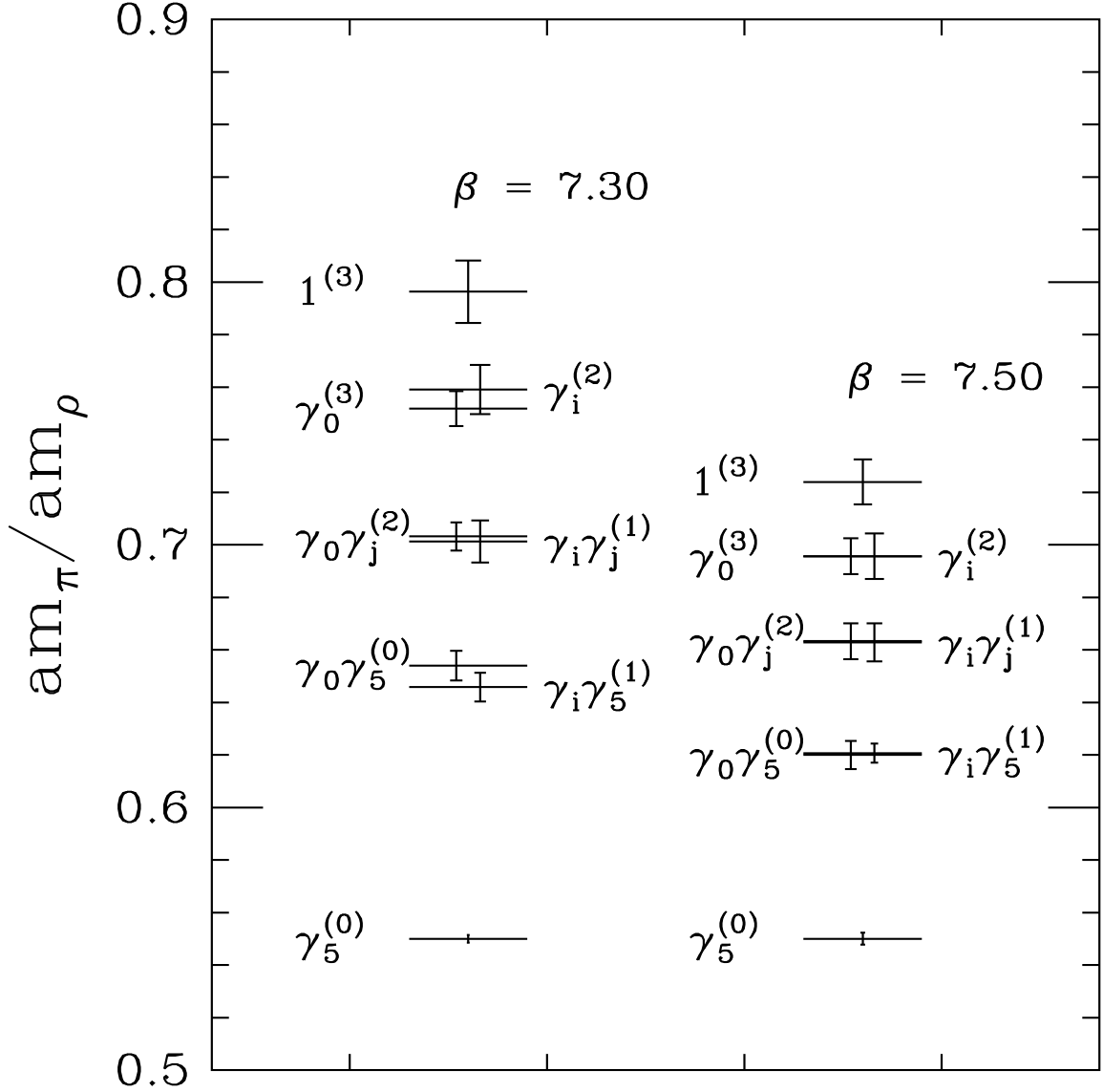}
\caption{
The spectrum of pion masses with the "Link+Staple+Naik" action.
The two spectra are at $10/g_{imp}^2=7.3$ and $7.5$ respectively.
In both cases we have interpolated in quark mass, assuming that squared
meson masses are linear in the quark mass, to the point where
$m_G/m_\rho=0.55$.
The pion masses are plotted in units of the $\rho$ mass.
Each pion is labeled by the gamma matrix specifying its flavor
structure.
The pattern of near degeneracies predicted in Ref.~\cite{SHARPE} is
evident.  As expected, the mass splittings among the pions are smaller
for the smaller lattice spacing ($10/g_{imp}^2=7.5$).
}
\label{TEMPLATE_OFN}
\end{figure}

\begin{figure}[tb]
\epsfxsize=6.5in
\epsfysize=3.0in
\epsfbox{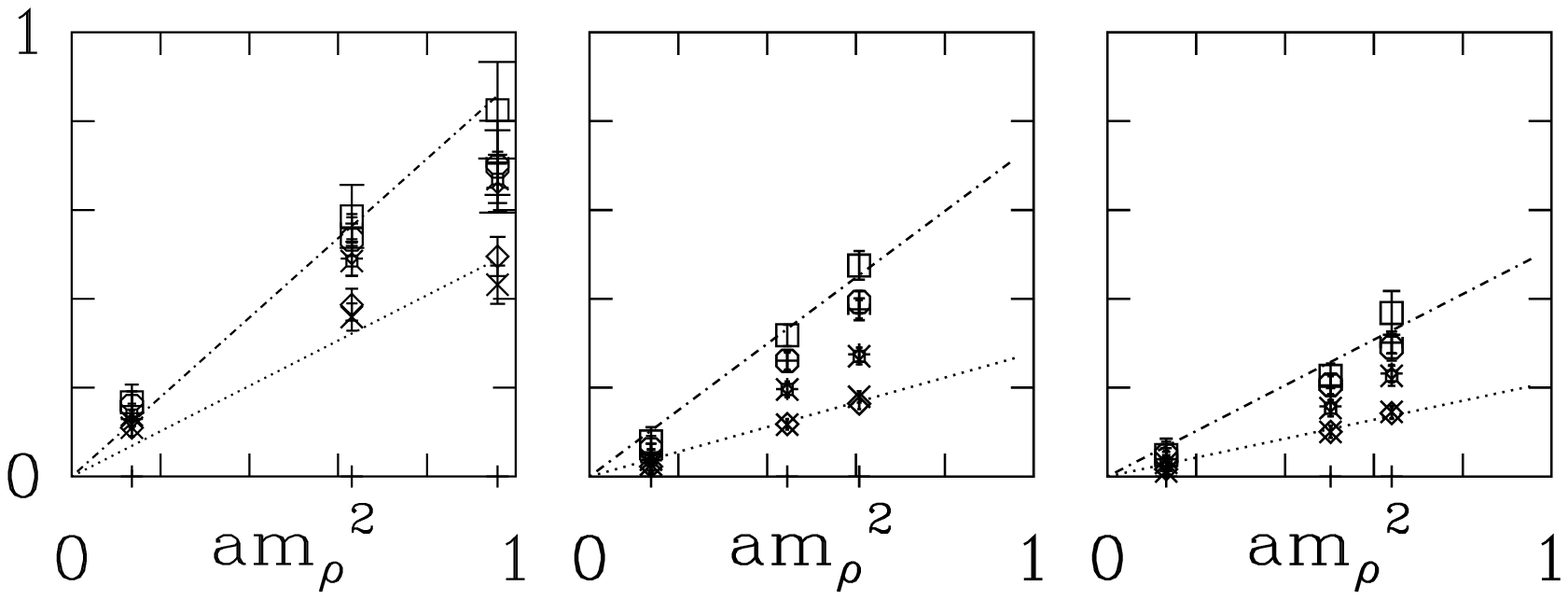}
\caption{
The $\delta_2$ for three different actions versus the squared lattice
spacing.  From left to right, the actions are the conventional ``One
Link'' action, the ``Link+Staple+Naik'' action, and the ``Order $a^2$
tadpole'' action.  In each case, the $\rho$ mass for the respective
action was used to define the length scale.  The points come from
two flavor runs with $10/g_{imp}^2=7.3$ (rightmost points in each
graph), $10/g_{imp}^2=7.5$, and a quenched simulation with single
plaquette gauge action at $6/g_{conv}^2=6.15$.  The lines drawn in each
panel are approximate slopes for the local non-Goldstone pion (lower line) and
for the three link flavor ${\bf 1}$ pion (upper line), which is the
worst case pion.
}
\label{THREE_SLOPES}
\end{figure}

Figure \ref{TEMPLATE_OFN}
shows the spectrum of pion masses obtained with
the ``Staple+Naik'' action at $10/g_{imp}^2=7.3$ and $7.5$
respectively.
Here, as in all of our calculations, the pattern of near-degeneracies predicted
in Ref. \cite{SHARPE} is evident.
Note that the local non-Goldstone pion, the flavor $\gamma_0 \gamma_5$
pion, is one of the lightest non-Goldstone pions, so a realistic
assesment of the flavor symmetry breaking requires consideration of the
nonlocal pions.
All to the actions we have tested produce a qualitatively similar
pattern of pion masses.

Figure \ref{THREE_SLOPES} shows $\delta_2$ from three different
actions, each with three different lattice spacings.  All of these
results are interpolated in quark mass to the point where
$m_G/m_\rho=0.55$.  The improvement of the flavor symmetry breaking
is evidenced by the smaller non-Goldstone pion masses for the improved
actions.  In this figure we have determined the lattice spacing for
each action using the $\rho$ mass evaluated with this action.  Thus,
even though the spectra for the different actions are evaluated on the
same sets of lattices, they appear at different horizontal positions.
However, as we would hope, the ambiguity in lattice spacing coming from
the difference in $\rho$ mass with action becomes smaller at the
smaller lattice spacings.
(It is neither surprising nor upsetting that the $\rho$ mass should
depend on the action.  After all, in the end we expect that these
actions will have better scaling behavior in all quantities, meaning
that we should come closer to the continuum limit at these large
lattice spacings.  Among other things, this means that we expect
the nucleon to rho mass ratio, a well known problem on coarse lattices,
to be improved with these actions.)


In Table \ref{DELTA_TABLE} we show $\delta_2$ for the various actions.
We tabulate this for only three of the pions: the local non-Goldstone
pion with flavor structure $\gamma_0 \gamma_5$, and the two three-link
pions, with flavor structures $\gamma_0$ and ${\bf 1}$.  The local
pion is the most often studied one, and so allows comparison with
other work.  The flavor ${\bf 1}$ is the worst case pion.  The other
three link pion is in the second worst multiplet, but it is
interesting to study because its parity partner would have exotic
quantum numbers, so the propagator can be fit to a simple exponential,
leading to smaller errors for the mass estimates.  We note in passing
that the flavor ${\bf 1}$ pion is properly called a ``pion'' here
instead of an ``eta'', since we did not compute quark-line
disconnected diagrams in the propagator.  In other words, one may
imagine that the quark and antiquark in this pion carry a flavor
quantum number in addition to that coming from the Kogut-Susskind
quarks' natural four flavors, and so cannot annihilate each other.
Thus, it is correct to demand that an improved action should make this
pion degenerate with the others.  In order to make a fair comparison
of the actions, we have interpolated (or extrapolated) the spectrum to
a fixed $m_G/m_{\rho}=0.55$ point. This is done using two different
bare masses for the valence quarks. We also interpolated (or
extrapolated) our
spectrum to a fixed $m_G=545MeV$ using the heavy quark potential 
parameter $r_0$ in
order to fix the scale\cite{SOMMER}. As we can see from the table, the computed
$\delta_2$'s in both cases are in agreement with in errors.  The
errors in both cases were computed with jackknife analysis.  Since
$\delta_2$ involves a ratio of mass differences, and the different
masses are all correlated, naive error propagation would lead to an
overestimate of the errors in $\delta_2$.  Therefore we used a
jackknife analysis to compute these errors.  When comparing actions,
we are interested in the difference in $\delta_2$ between the two
actions.  Since we used the same lattices for all the actions, these
$\delta_2$'s are not independent, and one should really do a jackknife
analysis in order to determine the errors on the differences of
$\delta_2$'s. We have done this and we find in general the same error
as the one computed with error propagation from the quoted $\delta_2$
errors.

Our results show that all the variants of the fat actions
significantly improve flavor symmetry, with larger improvement for the
actions that suppress more couplings to gluons with transverse
momentum $\pi$.  Generally, tadpole improving the tree level
coefficients results in a better action. The approximately 
reunitarized action with coefficient 0.25 seems to work better than
the Staple+Naik action. We have no clear understanding of why this
happens, but this may be a hint that a more careful 
(non-perturbative) tuning of the fattening coefficients would result
in  better actions.  As expected, the Fat7tad action is the best
for suppressing flavor symmetry violations. The Asqtad action is
slightly worse than the Fat7tad action in this regard.   Extra
$O(a^4)$ flavor symmetry violation introduced by the Naik term and the
Lepage term may be responsible for this.  However, the Asqtad action
has improved rotational symmetry due to the Naik term, and has no
additional $O(a^2)$ flavor conserving errors introduced by the
fattening.  Thus, of all the actions we have studied, we consider this
one to be the best candidate for an improved Kogut-Susskind action.

In Ref.~\cite{TD_AH_TK} it was shown that smearing the links by
``Ape smearing'', where the link and staples are averaged and the
result in projected back onto SU(3) (i.e. replaced by the SU(3) matrix
which maximizes $\Tr( U^\dagger F )$, where $F$ is the fattened link)
improved the flavor symmetry, among other nice features.
We find similar results on our set of sample lattices.  It is
interesting to compare the ``Ape1'' action in Table \ref{DELTA_TABLE} with the
``Staple-un(.25)'' action, since the difference between the two is that
the Ape1 action uses links that are exactly unitary, while the
Staple-un(.25) uses the same fattening, but is only unitary to first
order in the staple weight.  We see that the approximately unitary
action is only slightly worse than the Ape1 action.  Also, we see as
expected that that the Ape4 valence action gives a very good
suppression of flavor symmetry breaking, although using it for a
dynamical action is difficult.
The approximately reunitarized actions seem to have slightly better
flavor symmetry breaking than the comparable Link+Staple action, but 
probably not enough to justify the extra complexity.

Another practical advantage of the fat link actions is that the fat
link configurations are smoother than the original configurations, and
the conjugate gradient computation of the propagators converges in
fewer iterations.  Quantifying this statement is tricky because,
unlike the flavor symmetry breaking, where
$\delta_2$ is approximately independent of quark mass, the number of
conjugate gradient iterations is very sensitive to the quark mass.
For example, at $10/g_{imp}^2$ at a fixed bare quark mass of 0.02 the
Link+Staple action requires 21\% fewer conjugate gradient      
iterations than the conventional action.  However, after interpolation 
to $m_G/m_\rho=0.55$ this advantage disappears, and both of these
actions require about the same number of iterations.            
On the coarse lattice the approximately unitary actions do     
better --- the Staple-un(.25) action and the Ape1 action require about
10\% fewer iterations than the conventional or Link+Staple actions, and
the Ape4  21\% fewer.                                  
On the fine lattice, with $6/g_{conv}^2=6.15$, there is much less
ambiguity since the various pion masses are much closer.  Also,
the fattening does a much better job of smoothing the configurations  
on the fine lattice.  In this case the Link+Staple action requires 
46\% fewer iterations than the conventional action, and the Fat7 action 
52\% fewer.  However, the advantage of the approximately                
reunitarized action has disappeared -- it requires just about as many
iterations as the Link+Staple.  Regrettably, the Asqtad action does
not do as well as the Fat7 action, only reducing the number of iterations 
by 37\% as compared to                                            
the conventional action.  This is probably because of the negative   
coefficient associated with the Lepage term (see Table \ref{ACTIONTABLE}),
meaning that this term is actually undoing part of the smoothing
accomplished by the other paths.
We should note that with these complicated actions the cost of
simulation with dynamical quarks is no longer completely dominated
by the conjugate gradient.  Except for very light quark masses,
the cost of computing the fermion force and precomputing the fat links
becomes comparable to the conjugate gradient cost.

\begin{center}\begin{table}
\label{DELTA_TABLE}
\caption{
The flavor symmetry breaking measure $\delta_2$ for the various actions
and lattice spacings.  Results are shown for the local non-Goldstone
($\gamma_0\gamma_5$) pion and for the two three-link pions, with
Kogut-Susskind flavor structures $\gamma_0$ and ${\bf 1}$.
The actions are those specified in Table~\protect\ref{ACTIONTABLE},
plus actions with a single staple fattening approximately unitarized as
discussed in section 2.  For the actions with single staple fattening
we give the  relative weight of each staple to the single link in
parentheses ({\it e.g.} Staple-un(.25)).
The ``Ape1'' action is the single link plus the three link staple with
a relative weight of 0.25, projected back onto an element of SU(3) to
numerical precision.  For the ``Ape4'' action, this smearing and
projection was repeated four times.
Otherwise the coefficients of the various paths are
found in Table~\protect\ref{ACTIONTABLE}. For comparing the actions
we interpolate (or extrapolate) to {\protect$m_G/m_\rho = 0.55$} or  
alternatively to {\protect$m_G = 545 MeV$}.                          
}
\begin{tabular}{llll|lll}
& \multicolumn{3}{c}{$ m_G/m_\rho = 0.55 $} &               
   \multicolumn{3}{c}{$ m_G = 545 MeV $} \\                 
\hline
Action  &  $\gamma_0\gamma_5$ & $\gamma_0$ & ${\bf 1}$ 
        &  $\gamma_0\gamma_5$ & $\gamma_0$ & ${\bf 1}$ \\
\hline
\multicolumn{7}{l}{$10/g_{imp}^2 = 7.30$} \\
OneLink   &    0.408(45)  &    0.677(62)  &    0.815(73)  &    0.408(66)  &     
0.68(11)  &     0.83(15) \\
Staple+Naik   &   0.171(7)  &    0.370(18)  &    0.479(26)  &   0.170(7)  &  
  0.369(18)  &    0.479(25) \\
Ape1   &    0.141(13)  &    0.326(22)  &    0.424(22)  &    0.141(15)
&    0.326(27)  &    0.424(23) \\
Ape4   &    0.064(10)  &    0.209(10)  &    0.262(21)  &    0.070(36)
&    0.272(75)  &    0.296(80) \\
Staple-un(.25)   &   0.145(9)  &    0.320(31)  &    0.446(31)  &    0.150(17)  &  
  0.319(41)  &    0.457(51) \\
Fat5   &   0.126(8)  &    0.298(14)  &    0.393(16)  &   0.127(8)  &    0.299(14)  &    0.393(16) \\
Fat5tad   &   0.115(5)  &    0.277(18)  &    0.370(22)  &   0.116(5)  &    0.279(22)  &    0.375(28) \\
Fat7   &   0.117(9)  &    0.281(13)  &    0.367(14)  &   0.118(8)  &    0.282(13)  &    0.368(14) \\
Fat7tad   &   0.108(5)  &    0.260(16)  &    0.346(19)  &   0.109(6)  &    0.263(22)  &    0.353(28) \\
Asq   &    0.164(10)  &    0.357(21)  &    0.438(20)  &    0.164(12)  &    0.357(22)  &    0.439(20) \\
Asqtad   &   0.136(10)  &    0.279(21)  &    0.365(22)  &   0.137(10)  &    0.280(18)  &    0.367(20) \\
\hline
\multicolumn{7}{l}{$10/g_{imp}^2 = 7.50$} \\
OneLink   &    0.371(11)  &    0.569(13)  &    0.594(17)  &    0.363(18)  &    0
.526(22)  &    0.570(28) \\
Staple+Naik   &   0.115(5)  &   0.261(7)  &   0.318(8)  &   0.118(6)  &  
 0.266(8)  &   0.326(10) \\
Staple-un(.25)   &   0.103(8)  &    0.236(14)  &    0.304(20)  &    0.106(12)  &  
  0.241(21)  &    0.312(31) \\
Staple-un(.40)   &   0.131(8)  &    0.235(14)  &    0.288(19)  &    0.131(10)  &  
  0.236(18)  &    0.290(24) \\
Fat5   &   0.082(5)  &   0.200(7)  &   0.245(8)  &   0.086(5)  &   0.205(7)  &   0.256(8) \\
Fat5tad   &   0.080(6)  &    0.201(10)  &    0.261(13)  &   0.082(9)  &    0.206(16)  &    0.275(22) \\
Fat7   &   0.074(5)  &   0.187(7)  &   0.225(8)  &   0.076(5)  &   0.193(6)  &   0.236(8) \\
Fat7tad   &   0.071(7)  &    0.184(10)  &    0.236(13)  &   0.072(10)  &    0.189(17)  &    0.245(22) \\
Asq   &   0.115(5)  &   0.242(6)  &    0.288(11)  &   0.116(5)  &   0.246(8)  &    0.295(12) \\
Asqtad   &   0.101(7)  &    0.207(10)  &    0.255(11)  &   0.102(8)  &    0.207(12)  &    0.257(15) \\
\hline
\multicolumn{7}{l}{$6/g_{conv}^2 = 6.15$} \\
OneLink   &    0.096(38)  &    0.111(71)  &    0.157(41)  & ---  & ---  & --- \\
Staple+Naik   &    0.014(13)  &    0.050(19)  &    0.077(14)  & ---  & ---  & ---
 \\
Staple-un(.25)   &    0.026(27)  &    0.040(25)  &    0.057(24)  & ---  & ---  & ---
 \\
Fat5   &   0.011(9)  &    0.043(20)  &    0.057(23)  & ---  & ---  & --- \\
Fat7   &   0.009(8)  &    0.036(14)  &    0.049(24)  & ---  & ---  & --- \\
Asqtad   &    0.030(20)  &    0.034(18)  &    0.051(40)  & ---  & ---  & --- \\
\hline
\end{tabular}

\end{table}\end{center}

\section{Conclusions}

 In this paper we have investigated the possibility of constructing a
Kogut-Susskind action with improved flavor and rotational symmetry
suitable for dynamical fermion simulations. For this reason we want to
keep the amount of new paths introduced into the action as small as
possible.  An action containing fat links with paths up to length
seven was constructed. At tree level this action has no couplings of 
quarks to gluons with a transverse momentum component $\pi$. As a
result, at tree level the flavor symmetry violating terms in the
action are completely removed. In addition to flavor symmetry, the
rotational symmetry is improved by introducing the Naik term. Finally,
as Lepage pointed out\cite{LEPAGE98}, we need to introduce an    
extra five link staple in order to cancel errors of $O(a^2p^2)$
introduced by the fattening.  The resulting action can be further
improved by tadpole improvment.  This action, which we call
``Asqtad'', is an order $O(a^4,a^2g^2)$ accurate fermion action.

Asqtad is an action  simple enough to be useful for dynamical
simulations.  Preliminary tests using our generic code\cite{US_PAPER1} 
show about a factor of 4 higher cost than the code
implementing the standard Kogut-Susskind action. We have found
optimizations specific to the Asqtad action that could bring the cost
factor down to 2-2.5.  The $O(a^2)$ precision at a cost of a factor of
2-2.5 makes such an action very competitive with the other popular
improved actions such as D234\cite{D234}, perfect or approximately
perfect actions, the Neuberger action, and domain wall fermion
actions. The highly improved chiral symmetry of actions respecting  
(or approximately respecting) the Ginsparg-Wilson relation is not
something that Asqtad can compete with. On the other hand, cost
may favor the Asqtad action.
The Neuberger action\cite{URS,NEUBERGER}, approximately perfect   
actions\cite{DEGRAND98,TD_AH_TK,WB_HD}
and domain wall fermions\cite{BLUM} are all  
fairly costly to implement for dynamical fermions.

In view of the enormous price one has to pay in order to have highly
improved chiral symmetry on the lattice, we think that the Asqtad
action is a good candidate for a fermion action to be used in
the next generation of dynamical simulations. Flavor symmetry
breaking in the traditional Kogut-Suskind action at lattice spacings
commonly used in high temperature QCD studies results in
pions as heavy as the kaons, making it
impossible to study the effects of the strange quark. Our study of the
Asqtad action shows that one can achieve a good separation between the
pions and the kaons at accessible lattice spacings. Thus Asqtad is an
action that may prove very useful in projects in which the effects  
of the strange quark are to be studied.

\section*{Acknowledgements}
This work was supported by the U.S. Department of Energy under contract
DE-FG03-95ER-40906 
and by the National Science Foundation grant number
NSF--PHY97--22022. 
Computations were done on the
Paragon at Oak Ridge National
Laboratory, and the T3E's at NERSC, NPACI and the PSC.
We would like to thank Peter Lepage for helpful communications, and
the members of the MILC collaboration for inspiration and many
discussions.


\end{document}